\documentclass[twocolumn,preprintnumbers,10pt,prl]{revtex4-1}

\usepackage{graphicx}
\usepackage{amsmath}
\usepackage{amssymb}
\usepackage{verbatim}
\usepackage{mathrsfs}
\usepackage{array}
\usepackage{layout}
\usepackage{textcomp}
\usepackage{latexsym}
\usepackage{slashed}
\usepackage{rotating}
\usepackage{hyperref}
\usepackage{booktabs}
\usepackage{multirow}
\usepackage{relsize}
\usepackage[usenames,dvipsnames]{color}

\newcommand{\Eq}{&=&}

\newcommand{\nn}{\nonumber\\}
\newcommand{\hs}[1]{{\hspace{#1}}}
\newcommand{\vs}[1]{{\vspace{#1}}}
\newcommand{\tf}[1]{{\textsf{#1}}}

\begin{document}

\vspace*{-30mm}
\font\mini=cmr10 at 0.8pt

\title{Comment on $\tf{``}$New Freezeout Mechanism for Strongly Interacting Dark Matter$\tf{"}$}

\author{Shu-Yu\,\,Ho${}^{}$}\email{phyhunter@kias.re.kr}
%\author{Pyungwon\,\,Ko${}^{}$}\email{pko@kias.re.kr}
\author{Chih-Ting\,\,Lu${}^{}$}\email{timluyu@kias.re.kr\\}

\affiliation{\vspace{5pt} Korea Institute for Advanced Study, Seoul 02455, Republic of Korea} 

\preprint{KIAS-P21031}

\begin{abstract}
\end{abstract}

\maketitle

%%%%%%%%%%%%%%%%%%%%%%%%%%%%%%%%%%%%%%%%%%%%%%%%%

\begin{figure}[t!]
\begin{center}
\includegraphics[width=0.48\textwidth]{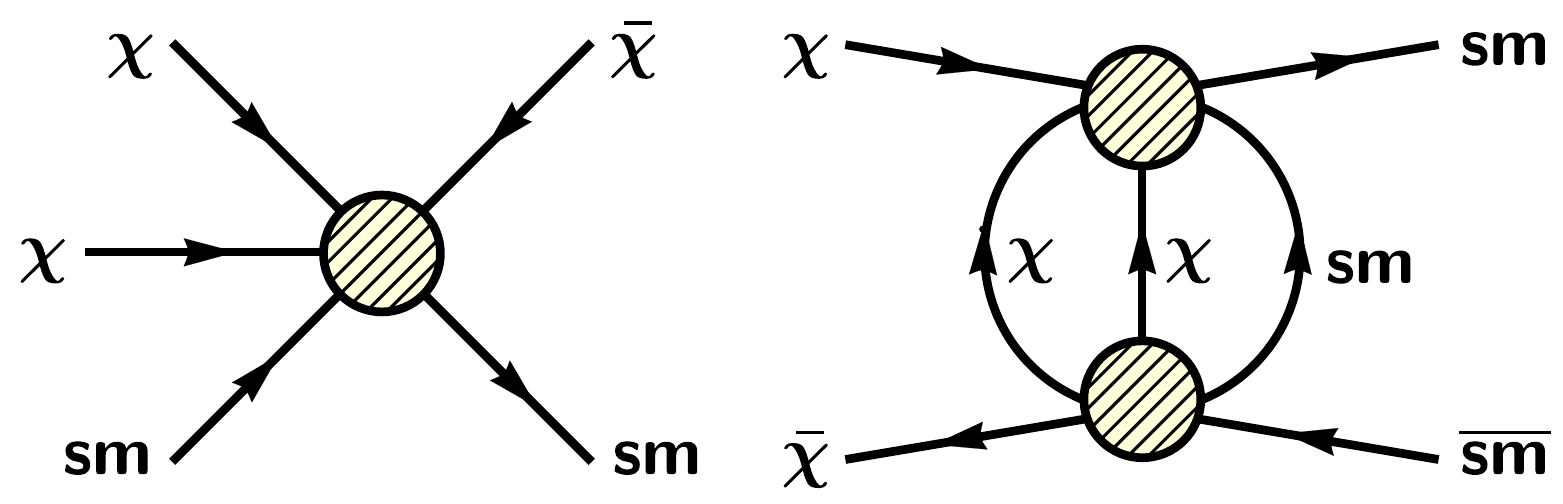}
\vs{-0.6cm}
\caption{Feynman diagrams of the $3 \to 2$ and $2 \to 2$ processes in the Co-SIMP paradigm.
\label{fig:CoSIMP}}
\end{center}
\vs{-0.2cm}
\end{figure}

In a recent Letter, J. Smirnov and J. F. Beacom \cite{Smirnov:2020zwf} proposed a new freezeout mechanism for strongly interacting dark matter (DM) dubbed Co-SIMP, where the reaction rate of the $3 \to 2$ process, $\chi + \chi + \tf{sm} \to \bar{\chi} + \tf{sm}$ with $\chi$ the Co-SIMP DM and $\tf{sm}$ the standard model (SM) particle, determines the relic abundance of DM\,; see the left graph in Fig.\,\ref{fig:CoSIMP}. In their work, they consider two cases for the Co-SIMP masses\,\,:\,\,(i) the typical case, $m^{}_\chi \ll m^{}_\tf{sm}$, and (ii) the edge case, $m^{}_\chi \simeq m^{}_\tf{sm}$, where $m^{}_\chi$ and $m^{}_\tf{sm}$ are the Co-SIMP and SM particle masses, respectively. Here we want to comment on case (ii).

As we have learned in our recent study \cite{Ho:2021ojb}, any five-point interaction attaching different DM species or SM particles can always generate
a two-loop $2 \to 2$ process that would enforce the heavier particles annihilate into the lighter particles\,; see the right graph of Fig.\,\ref{fig:CoSIMP}. Intuitively, one may think this process is suppressed by the two-loop factor and can be neglected in comparison to the $3 \to 2$ process. However, since the $3 \to 2$ process in the Co-SIMP scenario has to capture one extra nonrelativistic SM particle whose number yield is Boltzmann-suppressed, the reaction rate of the $2 \to 2$ process may dominate over or be comparable with that of the $3 \to 2$ process at the chemical freezeout of DM. Note that in Ref.\,\cite{Smirnov:2020zwf}, they do notice this two-loop diagram\,; however, what they concern about is the sensitivity of the elastic scattering cross section between DM and electron to the current and future direct detection experiments.

\begin{figure}[t!]
\begin{center}
\includegraphics[width=0.48\textwidth]{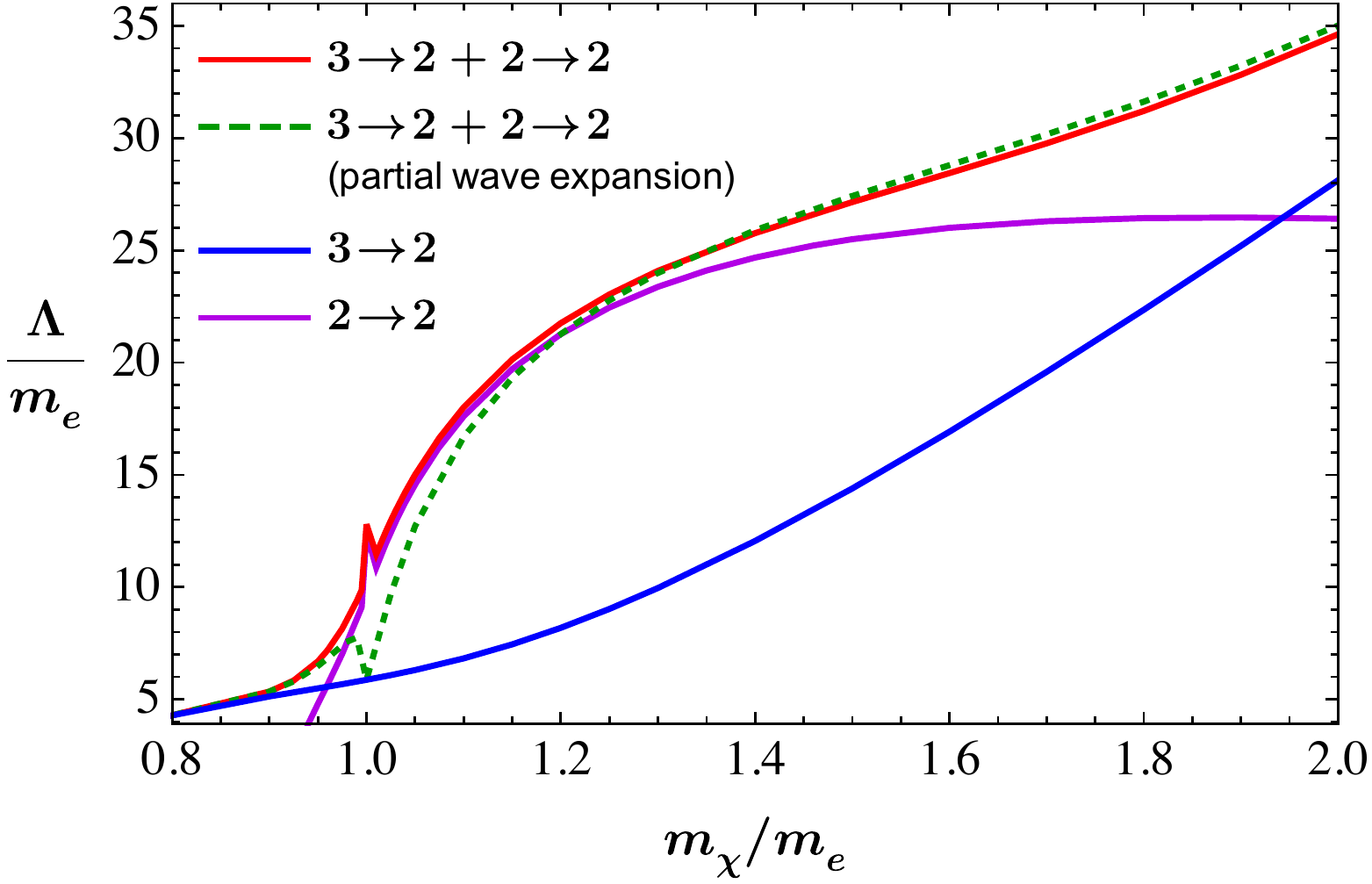}
\vs{-0.6cm}
\caption{Cutoff scale versus Co-SIMP mass in the edge case, where the red curve corresponds to the case with both $3 \to 2$ and $2 \to 2$ processes, and the blue (purple) curve corresponds to the case with only $3 \to 2$ ($2 \to 2$) processes, and the green dashed curve utilizes the $2 \to 2$ annihilation cross sections in partial wave expansion.
\label{fig:Lambda_mchi}}
\end{center}
\vs{-0.5cm}
\end{figure}

To illustrate our point explicitly, here we consider an electrophilic model in Ref.\,\cite{Smirnov:2020zwf}, where the Co-SIMP DM couples to the electron field, $e{}^{}$. The effective operators describing the $3 \to 2$ and two-loop induced $2 \to 2$ processes are given by
\begin{eqnarray}\label{int}
{\cal O}^{}_{3 \to 2}
\,=\,
\frac{1}{3!\Lambda^2} \chi^3 \bar{e} {}^{}e 
~,\quad
{\cal O}^{}_{2 \to 2}
\,=\,
\frac{c^{}_{\chi e}}{\Lambda} \bar{\chi} \chi \bar{e} {}^{} e
~, 
%\nonumber
\end{eqnarray}
where $\Lambda$ being the cutoff scale, and $c^{}_{\chi e}$ is the dimensionless loop-induced coupling of the form computed in Ref.\,\cite{Smirnov:2020zwf}.\,\,With these interactions, we then calculate the annihilation cross sections for all possible $3 \to 2$ processes, $\chi\chi\chi \to e^+e^-, \chi\chi e^{\pm} \to \bar{\chi} e^{\pm}$ and $\chi e^+e^- \to \bar{\chi}\bar{\chi}{}^{}$, as well as $2 \to 2$ processes, $\chi\bar{\chi} \to e^+e^-$ and $e^+e^- \to \chi\bar{\chi}{}^{}$, and derive the Boltzmann equation of the number density for $\chi$. We summarize them in the appendix. After numerically solving the Boltzmann equation, we obtain the result as shown in Fig.\,\ref{fig:Lambda_mchi}, where color curves satisfy the observed DM relic density. As indicated, the cutoff scale is enhanced by a factor of $2 \sim 3$ depending on the Co-SIMP mass. Therefore, the $2 \to 2$ process does affect the thermal history of the Co-SIMP DM for the edge case. In particular, we find that the reaction rate of the $2 \to 2$ process is dominated in the mass range $1 \lesssim m^{}_\chi/m^{}_e \lesssim 1.6$, and becomes subdominant in the mass range $1.6 \lesssim m^{}_\chi/m^{}_e \lesssim 2$. In addition, our numerical calculation shows that the freezeout temperature of the Co-SIMP DM is around $x^{}_\tf{f.o.} \hs{-0.05cm} \simeq 14 \sim 16$ which is bigger than the estimate in Ref.\,\cite{Smirnov:2020zwf}, where $x^{}_\tf{f.o.} \hs{-0.05cm} \simeq 10$\,; see the figures in the appendix.

Finally, let us point out that the relative interaction strengths of $3 \to 2$ and $2 \to 2$ processes may depend on UV completion models.\,\,One just keeps in mind that the $2 \to 2$ processes may give some effects on the Co-SIMP mechanism, especially for the edge case. 


The authors want to thank Pyungwon Ko for discussion. This work is supported by KIAS Individual Grants under Grant No.\,PG081201 (S.Y.H.), and No.\,PG075301 (C.T.L.).

\section{Appendix}
\vs{-0.3cm}
The full Boltzmann equation of the comoving number density $Y^{}_\chi$ for the Co-SIMP, $\chi{}^{}$, including the $3 \to 2 $ and $2 \to 2$ processes and assuming no asymmetry between $\chi$ and $\bar{\chi}$ in the electrophilic model is given by
\begin{eqnarray}\label{BEQ}
\hs{-0.4cm}
\frac{\text{d}Y^{}_\chi}{\text{d}x} 
\Eq
-{}^{}\frac{s^2}{H x}
\scalebox{1.1}{\bigg\{}
12{}^{}{}^{}\langle \sigma v^2 \rangle^{}_{\hs{-0.03cm}\chi\chi\chi \to e^+ e^-}
\scalebox{1.2}{\big[} Y_\chi^3 - \big(Y^\tf{eq}_\chi\big){}^{\hs{-0.03cm}3} \scalebox{1.2}{\big]}
\nn
&&\hs{1.15cm}
{+}\,2{}^{}{}^{}\langle  \sigma v^2 \rangle^{}_{\hs{-0.03cm}\chi\chi e^+ \to \bar{\chi} e^+} {}^{}
Y^{}_\chi Y^\tf{eq}_e \scalebox{1.2}{\big(} Y^{}_\chi - Y^\tf{eq}_\chi \scalebox{1.2}{\big)}
\nn
&&\hs{1.15cm}
{+}\,2{}^{}{}^{}\langle  \sigma v^2 \rangle^{}_{\hs{-0.03cm}\chi\chi e^- \to \bar{\chi} e^-} {}^{}
Y^{}_\chi Y^\tf{eq}_e \scalebox{1.2}{\big(} Y^{}_\chi - Y^\tf{eq}_\chi \scalebox{1.2}{\big)}
\nn
&&\hs{1.15cm}
{-}{}^{}{}^{}\langle  \sigma v^2 \rangle^{}_{\hs{-0.03cm}\chi e^+ e^- \to \bar{\chi} \bar{\chi}} {}^{}{}^{}
Y^{}_\chi \big(Y^\tf{eq}_e\big){}^{\hs{-0.03cm}2} \bigg( 1 - \frac{Y^{}_\chi}{Y^\tf{eq}_\chi} \bigg)
\hs{-0.05cm}
\scalebox{1.1}{\bigg\}}
\nonumber\\
&&
-{}^{}\frac{s}{H x}
\scalebox{1.1}{\bigg\{}
4{}^{} \langle \sigma v \rangle^{}_{\hs{-0.03cm}\chi\bar{\chi} \to e^+ e^-} 
\scalebox{1.2}{\big[} Y_\chi^2 - \big(Y^\tf{eq}_\chi\big){}^{\hs{-0.03cm}2} \scalebox{1.2}{\big]}
\nn
&&\hs{1.15cm}
{-}{}^{}{}^{}\langle \sigma v \rangle^{}_{\hs{-0.03cm}e^+ e^- \to \chi\bar{\chi}} \,
\big(Y^\tf{eq}_e\big){}^{\hs{-0.03cm}2} 
\scalebox{1.1}{\bigg[} 1- \frac{Y_\chi^2}{\big(Y^\tf{eq}_\chi\big){}^{\hs{-0.03cm}2}}
\scalebox{1.1}{\bigg]}
\hs{-0.05cm}
\scalebox{1.1}{\bigg\}}
\,,
\end{eqnarray}
where $x \equiv m^{}_\chi/T$ is the dimensionless time variable, $Y^\tf{eq}_i$ is the equilibrium comoving number yield of the species $i$ with the internal degrees of freedom $g^{}_i{}^{}$\,\cite{Gondolo:1990dk},
\begin{eqnarray}
Y^\tf{eq}_i
\,=\,
\frac{45}{4\pi^4}
\frac{g^{}_i}{g^{}_{s\ast}(x)}
\bigg(\frac{m^{}_i {}^{} x}{m^{}_\chi}\bigg)^{\hs{-0.13cm}2} K^{}_2\bigg(\frac{m^{}_i {}^{} x}{m^{}_\chi}\bigg)
~,
\end{eqnarray}
and $s$ and $H$ are the comoving entropy density and the Hubble parameter, respectively, given by
\begin{eqnarray}
s = \frac{2{}^{}\pi^2}{45} g^{}_{\star s}(x) {}^{}{}^{} \frac{m_\chi^3}{x^3} ~,\quad
H = \sqrt{\frac{\pi^2 g^{}_{\star}(x)}{90}} \frac{m_\chi^2}{x^2 m^{}_\tf{Pl}} 
\end{eqnarray}
with $g^{}_{\star}\,(g^{}_{\star s})$ being the effective energy (entropy) degrees of freedom of the thermal bath~\cite{Saikawa:2018rcs} and $m^{}_\tf{Pl}$ the reduced Planck mass. Using the interactions in Eq.\,\eqref{int}, the $3 \to 2$ annihilation cross sections are calculated as
\begin{eqnarray}
\langle \sigma v^2 \rangle^{}_{\hs{-0.03cm} \chi\chi\chi \to e^+e^-}
\Eq
\frac{3}{256{}^{}\pi{}^{}m^{}_\chi \Lambda^4} 
\scalebox{1.1}{\bigg(} \hs{-0.03cm}
1 - \frac{4{}^{}m_e^2}{9{}^{}m_\chi^2}
\scalebox{1.1}{\bigg)}^{\hs{-0.13cm}3/2}
~\,,\quad
\nn[0.12cm]
\langle \sigma v^2 \rangle^{}_{\hs{-0.03cm} \chi\chi e^\pm \to \bar{\chi} e^\pm}
\Eq
\frac{\sqrt{3}}{128{}^{}\pi{}^{}m^{}_\chi \Lambda^4} 
\frac{m_\chi^2 + 2{}^{} m^{}_\chi m^{}_e + 2{}^{}m_e^2}
{\big(m^{}_\chi + m^{}_e \big) \big(2{}^{}m^{}_\chi + m^{}_e \big)\raisebox{1pt}{$^{\hs{-0.04cm}2}$}}
\nn
&&\times
\sqrt{3{}^{}m_\chi^2 + 8{}^{} m^{}_\chi m^{}_e + 4{}^{}m_e^2}  
~\,,\quad
\nn[0.1cm]
\langle \sigma v^2 \rangle_{\hs{-0.03cm}\chi e^+e^- \to \bar{\chi}\bar{\chi}} \Eq {\cal O}\big(x^{-1}\big)
~,
\end{eqnarray}
here we have used the cross section formula of the $3 \to 2$ process, $1+2+3 \to 4+5$, in the non-relativistic limit as
\begin{eqnarray}
\big(\sigma v^2\big)_{\hs{-0.03cm}123 \to 45}
&{}^{}{}^{}\approx{}^{}{}^{}&
\frac{
\sqrt{{\cal K}\hs{-0.03cm}\scalebox{1.2}{\big[}\big(m^{}_1 + m^{}_2 + m^{}_3\big)\raisebox{0.5pt}{$^{\hs{-0.03cm}2}$}, m_4^2, m_5^2 \scalebox{1.2}{\big]}}}
{64{}^{} \pi {}^{} m^{}_1m^{}_2 m^{}_3  \big(m^{}_1 + m^{}_2 + m^{}_3\big)\raisebox{0.5pt}{$^{\hs{-0.03cm}2}$}} 
\label{cs3t2}
\nn
&&\times
\overline{\big|{\cal M}^{}_{123 \to 45}\big|\raisebox{0.5pt}{$^{\hs{-0.01cm}2}$}} ~,
\\[0.1cm]
{\cal K}\big(a,b,c\big) 
\Eq a^2+b^2+c^2-2\big(a{}^{}b+b{}^{}c+a{}^{}c\big) ~.
\end{eqnarray}
Notice that the matrix element squared in Eq.\,\eqref{cs3t2} is the one defining in the Boltzmann equation with averaging over initial and final spins, and including appropriate symmetry factors for identical particles in the initial or final states~\cite{Kolb:1990vq}.\,\,On the other hand, we employ the following Mandelstam variables in the nonrelativistic limit to evaluate the matrix element squared,
\begin{eqnarray}
\hs{-0.5cm}
s^{}_{jk} \Eq \big({}^{}{}^{}p^{}_j + p^{}_k\big)\raisebox{1pt}{$^{\hs{-0.03cm}2}$} \approx \big(m^{}_j + m^{}_k {}^{} \big)\raisebox{1pt}{$^{\hs{-0.03cm}2}$}
~,
\\[0.1cm]
\hs{-0.5cm}
s^{}_{45} \Eq \big({}^{}{}^{}p^{}_4 + p^{}_5\big)\raisebox{1pt}{$^{\hs{-0.03cm}2}$} \approx \big(m^{}_1 + m^{}_2 + m^{}_3 {}^{} \big)\raisebox{1pt}{$^{\hs{-0.03cm}2}$}
~,
\\[0.1cm]
\hs{-0.5cm}
t^{}_{k\ell} \Eq \big({}^{}{}^{}p^{}_k - p^{}_\ell\big)\raisebox{1pt}{$^{\hs{-0.03cm}2}$}
\approx \big(m^{}_k - m^{}_\ell {}^{} \big)\raisebox{1pt}{$^{\hs{-0.03cm}2}$}
- \frac{2{}^{}m^{}_k{}^{}\mu^{}_{45} \Delta m}{m^{}_\ell}
\end{eqnarray}
with $j,k = \{1,2,3\},\,\ell=\{4,5\}$ and
\begin{eqnarray}
\mu^{}_{45}
{}^{}\Eq{}^{}
\frac{m^{}_4 m^{}_5}{m^{}_4 + m^{}_5}
~,
\\[0.1cm]
\Delta m 
{}^{}\Eq{}^{}m^{}_1 + m^{}_2 + m^{}_3 - m^{}_4 - m^{}_5
~,
\end{eqnarray}
and these Mandelstam variables satisfy the relation
\begin{eqnarray}
&&s^{}_{12} + s^{}_{13} + s^{}_{23} + s^{}_{45}
\nn
&&+\,{}^{}t^{}_{14} + t^{}_{24} + t^{}_{34} + t^{}_{15} + t^{}_{25} + t^{}_{35}
\nonumber\\[0.12cm]
\Eq {}^{} 3\big(m_1^2 + m_2^2 + m_3^2 + m_4^2 + m_5^2 {}^{}\big)
~.
\end{eqnarray}
Also, the thermally-averaged $2 \to 2$ annihilation cross sections are calculated as~\cite{Gondolo:1990dk}
\begin{eqnarray}
\hs{-0.5cm}
\langle \sigma v \rangle^{}_{\hs{-0.03cm}\chi \bar{\chi} \to e^+e^-}
\Eq
\frac{x}{8{}^{}m_\chi^5 \big[K^{}_2(x)\big]\raisebox{1pt}{$^{\hs{-0.03cm}2}$}}
\nn
&&
{\times}
\mathop{\mathlarger{\int}_{\hs{-0.03cm}4m_\chi^2}^\infty} \hs{-0.02cm} \text{d}s^{}_\chi
{}^{}{}^{} \sqrt{s^{}_\chi} {}^{}{}^{} \scalebox{1.1}{\big(} s^{}_\chi - 4{}^{}m_\chi^2 {}^{}\scalebox{1.1}{\big)} {}^{}{}^{}
\nn[-0.12cm]
&&\hs{1.15cm}
{\times}{}^{}{}^{}
K^{}_1 \hs{-0.03cm}
\bigg(\hs{-0.05cm}
\frac{\sqrt{s^{}_\chi} {}^{}{}^{} x}{m^{}_\chi}
\bigg){}^{}{}^{}
\sigma^{}_{\chi \bar{\chi} \to e^+e^-}
~,
\\[0.1cm]
\hs{-0.5cm}
\langle \sigma v \rangle^{}_{\hs{-0.03cm}e^+e^- \to \chi \bar{\chi}}
\Eq
\frac{x}{8{}^{}m_e^4 {}^{} m^{}_\chi \big[K^{}_2(m^{}_e {}^{} x / m^{}_\chi)\big]\raisebox{1pt}{$^{\hs{-0.03cm}2}$}}
\nn
&&
{\times}
\mathop{\mathlarger{\int}_{\hs{-0.03cm}4m_e^2}^\infty} \hs{-0.02cm} \text{d}s^{}_e
{}^{}{}^{} \sqrt{s^{}_e} {}^{}{}^{} \scalebox{1.1}{\big(} s^{}_e - 4{}^{}m_e^2 {}^{}\scalebox{1.1}{\big)} {}^{}{}^{}
\nn[-0.12cm]
&&\hs{1.15cm}
{\times}{}^{}{}^{}
K^{}_1 \hs{-0.03cm}
\bigg(\hs{-0.05cm}
\frac{\sqrt{s^{}_e} {}^{}{}^{} x}{m^{}_\chi}
\bigg){}^{}{}^{}
\sigma^{}_{e^+e^- \to \chi \bar{\chi}}
~,
\end{eqnarray}
where $s^{}_\chi = \big({}^{}{}^{}p^{}_\chi + p^{}_{\bar{\chi}}{}^{}\big)\raisebox{0.1pt}{$^{\hs{-0.03cm}2}$}$ and $s^{}_e = \big({}^{}{}^{}p^{}_{e^+} + p^{}_{e^-}\hs{-0.03cm}\big)\raisebox{0.1pt}{$^{\hs{-0.03cm}2}$}$, and
\begin{eqnarray}
\hs{-0.5cm}
\sigma^{}_{\chi \bar{\chi} \to e^+e^-}
\Eq
\frac{c_{\chi e}^2}
{32{}^{}\pi{}^{}\Lambda^2}
\frac{\big(s^{}_\chi - 4{}^{}m_e^2\big)\raisebox{1pt}{$^{\hs{-0.03cm}3/2}$}}
{s^{}_\chi \big(s^{}_\chi - 4{}^{}m_\chi^2\big)\raisebox{1pt}{$^{\hs{-0.03cm}1/2}$}}
~,
\\[0.1cm]
\hs{-0.5cm}
\sigma^{}_{e^+e^- \to \chi \bar{\chi}}
\Eq
\frac{c_{\chi e}^2}
{32{}^{}\pi{}^{}\Lambda^2}
\frac{\big(s^{}_e - 4{}^{}m_\chi^2\big)\raisebox{1pt}{$^{\hs{-0.03cm}1/2}$}
\big(s^{}_e - 4{}^{}m_e^2\big)\raisebox{1pt}{$^{\hs{-0.03cm}1/2}$}}
{s^{}_e}
\end{eqnarray}
with~\cite{Smirnov:2020zwf}
\begin{eqnarray}
c^{}_{\chi e}
\,\approx\,
\frac{m^{}_e}{(4\pi)^4 \Lambda}
\scalebox{1.05}{\bigg(} 1-\frac{m_\chi^2}{\Lambda^2} \scalebox{1.05}{\bigg)}
\ln \hs{-0.05cm} \scalebox{1.05}{\bigg(} \frac{\Lambda^2 + m_e^2}{4{}^{}m_\chi^2} \scalebox{1.05}{\bigg)}
~.
\end{eqnarray}
Finally, the prefactor of each cross section in Eq.\,\eqref{BEQ} is the product of the number difference of $\chi/\bar{\chi}$ in the initial and final states and the internal degrees of freedom in the final states~\cite{Ho:2021ojb}.

\begin{figure}[t!]
\begin{center}
\includegraphics[width=0.48\textwidth]{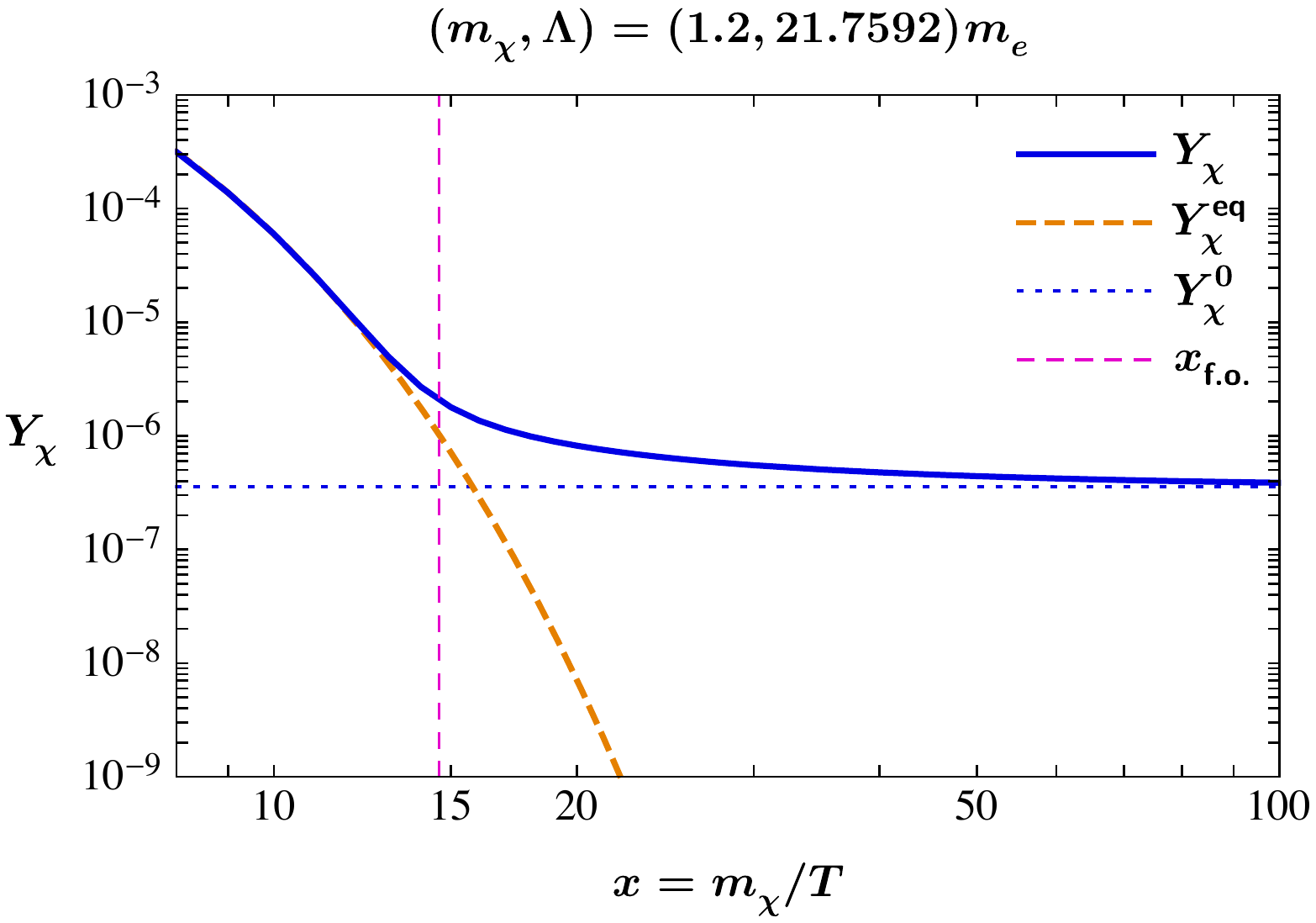}
\\[0.2cm]
\includegraphics[width=0.48\textwidth]{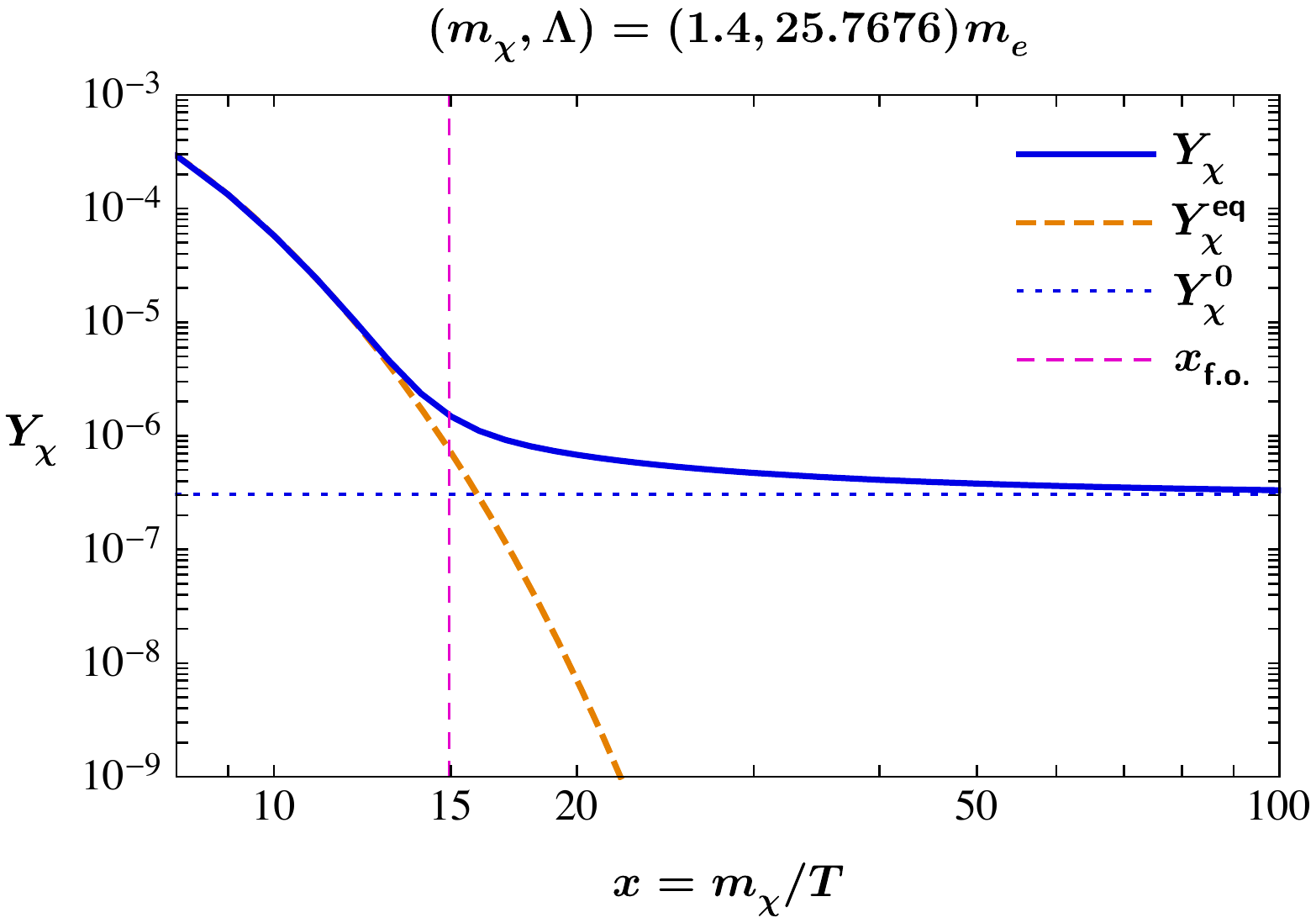}
\vs{-0.6cm}
\caption{$Y^{}_\chi(x)$ for $m^{}_\chi = 1.2{}^{}{}^{}m^{}_e$ and $1.4{}^{}{}^{}m^{}_e$.
\label{fig:Ychi1}}
\end{center}
\vs{-0.48cm}
\end{figure}

\begin{figure}[t!]
\begin{center}
\includegraphics[width=0.48\textwidth]{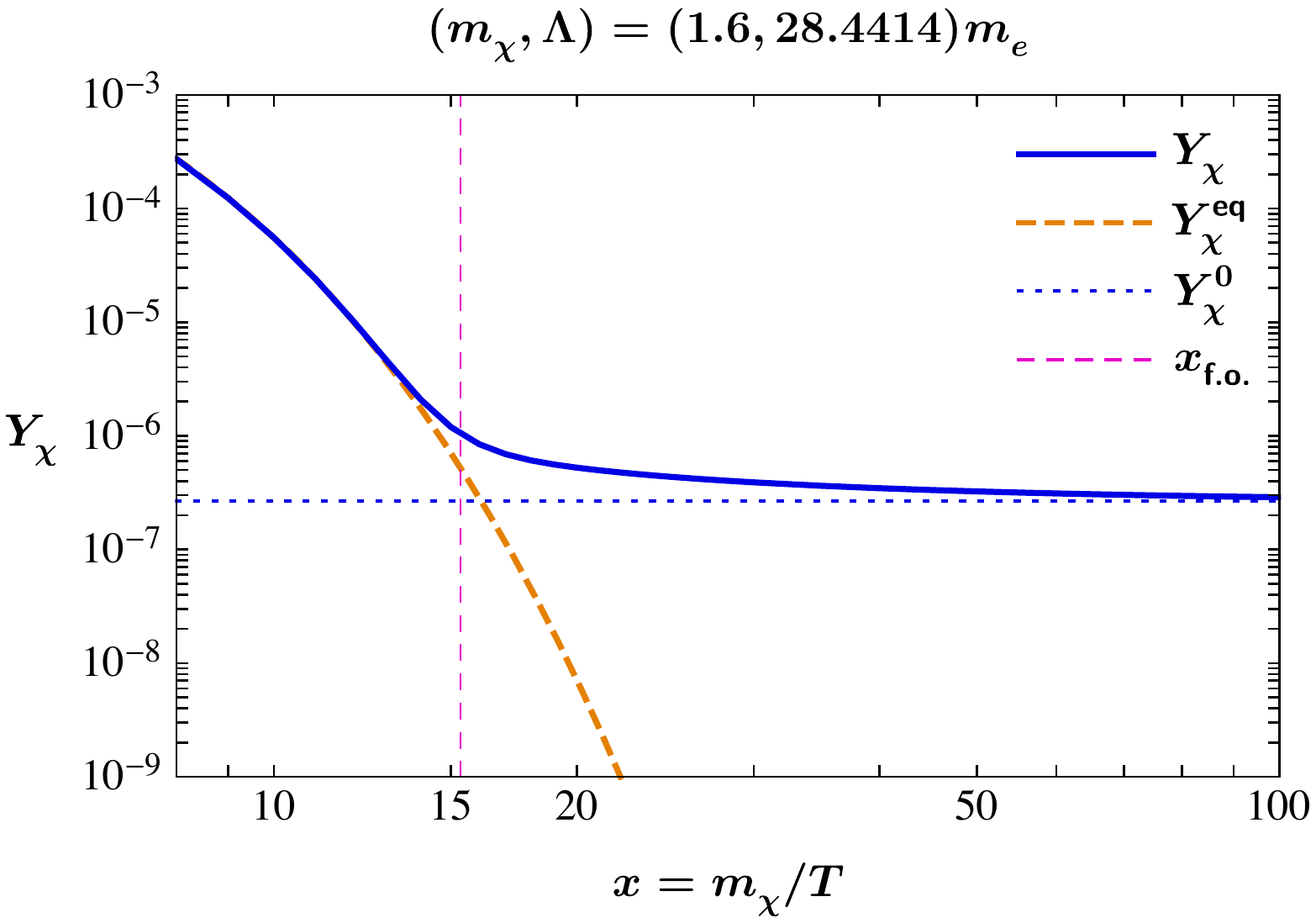}
\\[0.2cm]
\includegraphics[width=0.48\textwidth]{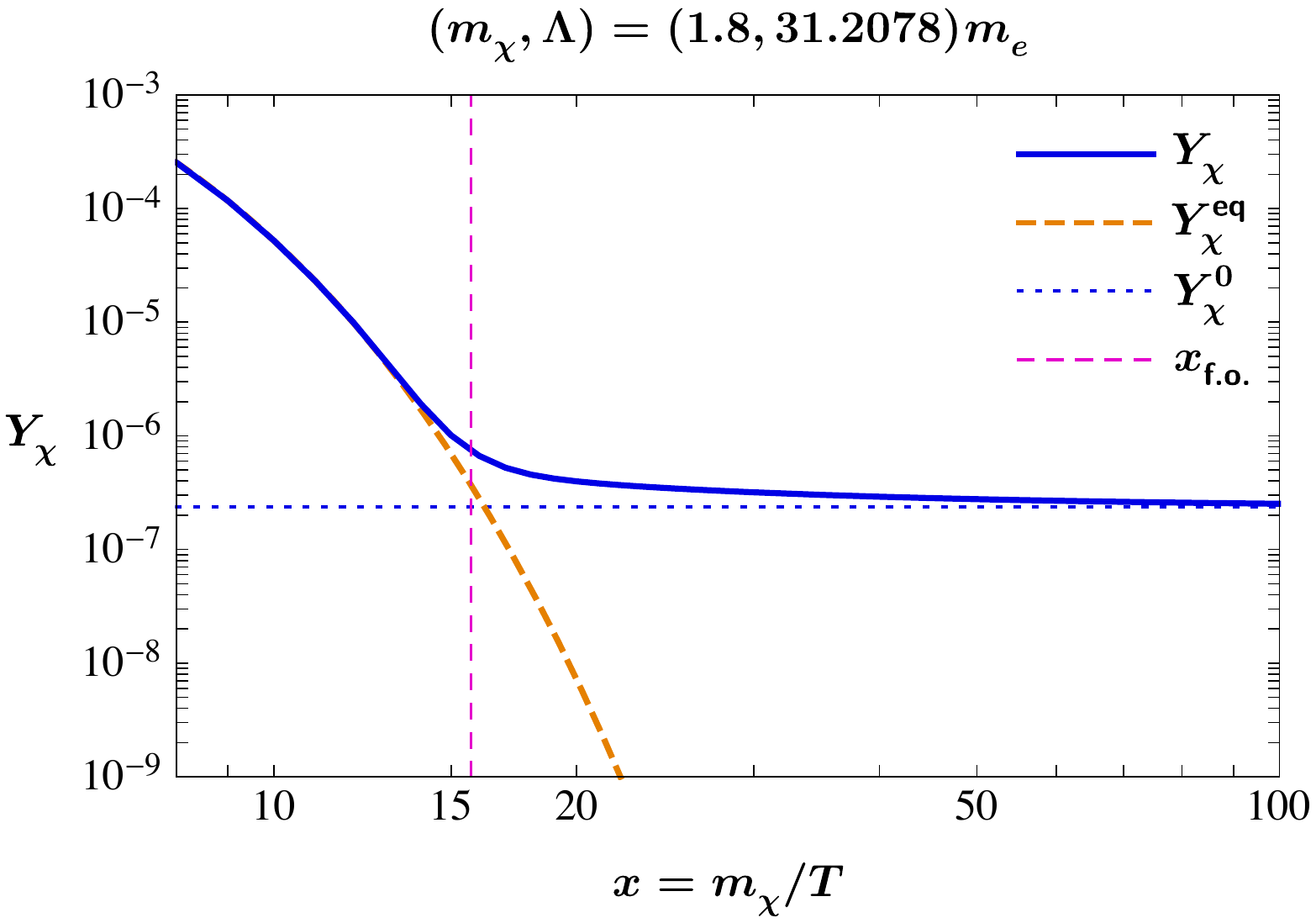}
\vs{-0.6cm}
\caption{$Y^{}_\chi(x)$ for $m^{}_\chi = 1.6{}^{}{}^{}m^{}_e$ and $1.8{}^{}{}^{}m^{}_e$.
\label{fig:Ychi2}}
\end{center}
\vs{-1.54cm}
\end{figure}

Now, solving Eq.\,\eqref{BEQ} with the proper initial condition $Y^{}_\chi \big(3 \lesssim x^{}_\tf{ini.} \hs{-0.06cm} \ll \hs{-0.03cm} x^{}_\tf{f.o.} \big) = Y^\tf{eq}_\chi(x^{}_\tf{ini.})$, we can obtain the cosmological evolution of the comoving number yield of $\chi$ as a function of $x$, $Y_\chi(x)$, and then predict the present density of $\chi$ by the relation below\,\cite{Bhattacharya:2019mmy}
\begin{eqnarray}
\Omega_\chi h^2 
\,\simeq\,
5.49 \times 10^5 {}^{}
Y^0_\chi
\scalebox{0.9}{\bigg(}
\frac{m^{}_\chi}{\text{MeV}}
\scalebox{0.9}{\bigg)}
~,
\end{eqnarray}
where $Y^0_\chi = Y^{}_\chi({}^{} x\to\infty)$. Imposing the observed DM abundance, $\Omega^{}_\tf{DM} h^2 = 0.12 \pm 0.0012$\,\cite{Aghanim:2018eyx}, one can fix the value of  $\Lambda$ for a given $m^{}_\chi$.\,\,We display in Figs.\,\ref{fig:Ychi1} and \ref{fig:Ychi2} a few examples of $Y^{}_\chi(x)$ in the mass range of interest, where the $\Lambda$ values are tunned to fit $\Omega^{}_\tf{DM} h^2 = 0.12$. As we can see from these figures, the freezeout temperature of the Co-SIMP DM is about $x^{}_\tf{f.o.} \simeq 14 \sim 16$, where we define the $x^{}_\tf{f.o.}$ which satisfies the condition $\Delta (x^{}_\tf{f.o.}) = Y^\tf{eq}_\chi(x^{}_\tf{f.o.})$ with $\Delta (x) = Y^{}_\chi(x)- Y^\tf{eq}_\chi(x)$\,\cite{Kolb:1990vq}.

\begin{figure}[t!]
\begin{center}
\includegraphics[width=0.46\textwidth]{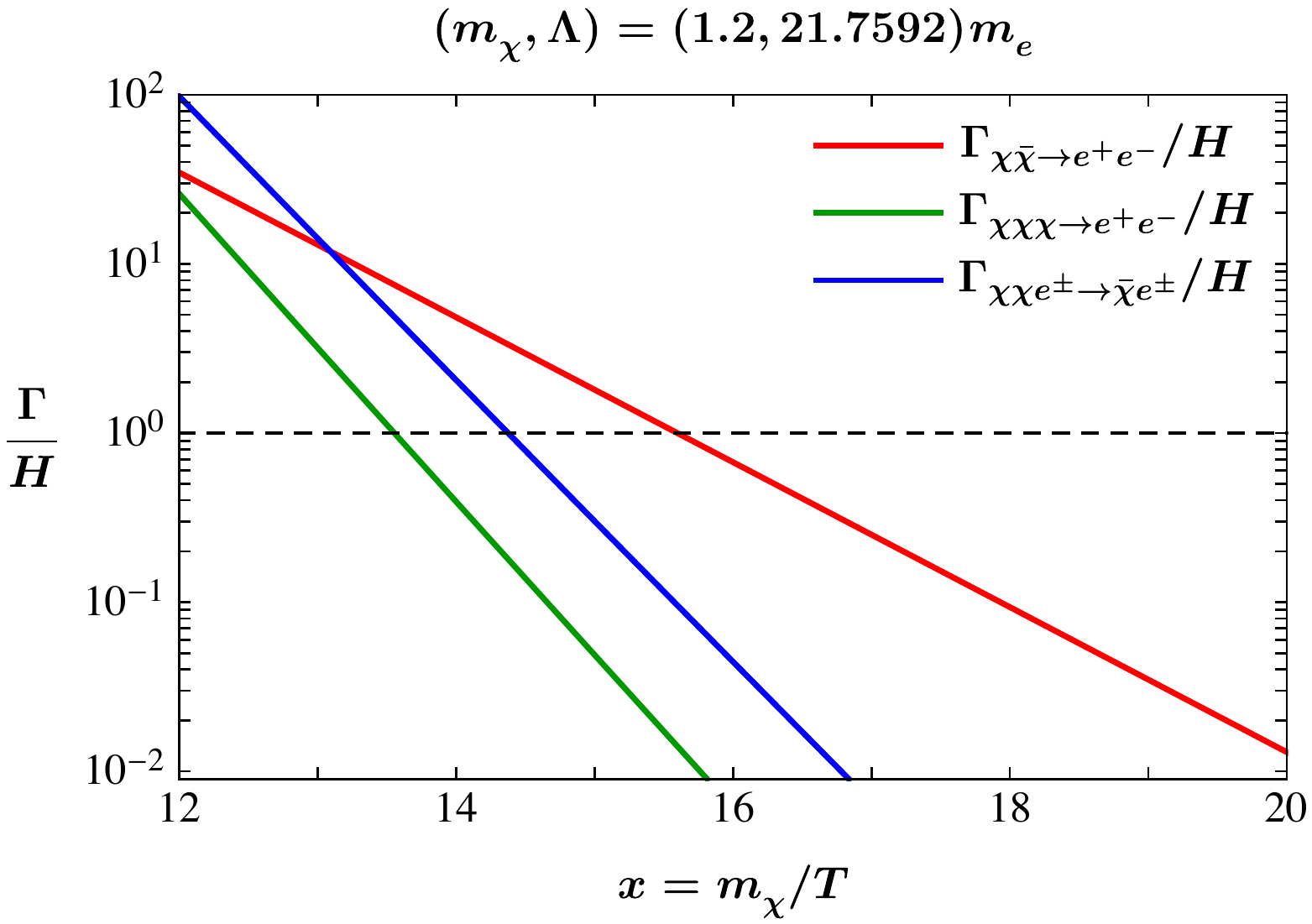}
\\[0.2cm]
\includegraphics[width=0.46\textwidth]{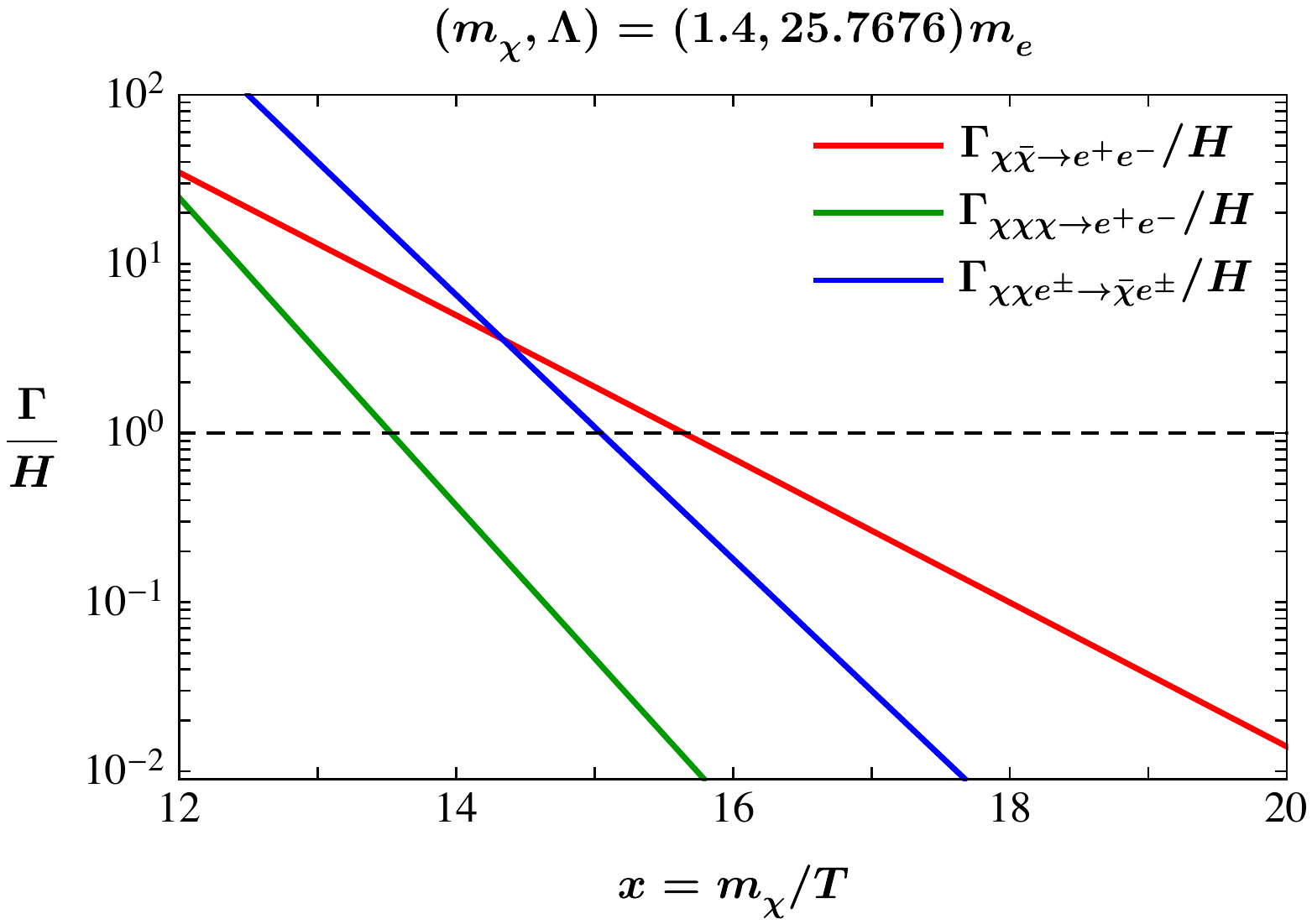}
\\[0.2cm]
\includegraphics[width=0.46\textwidth]{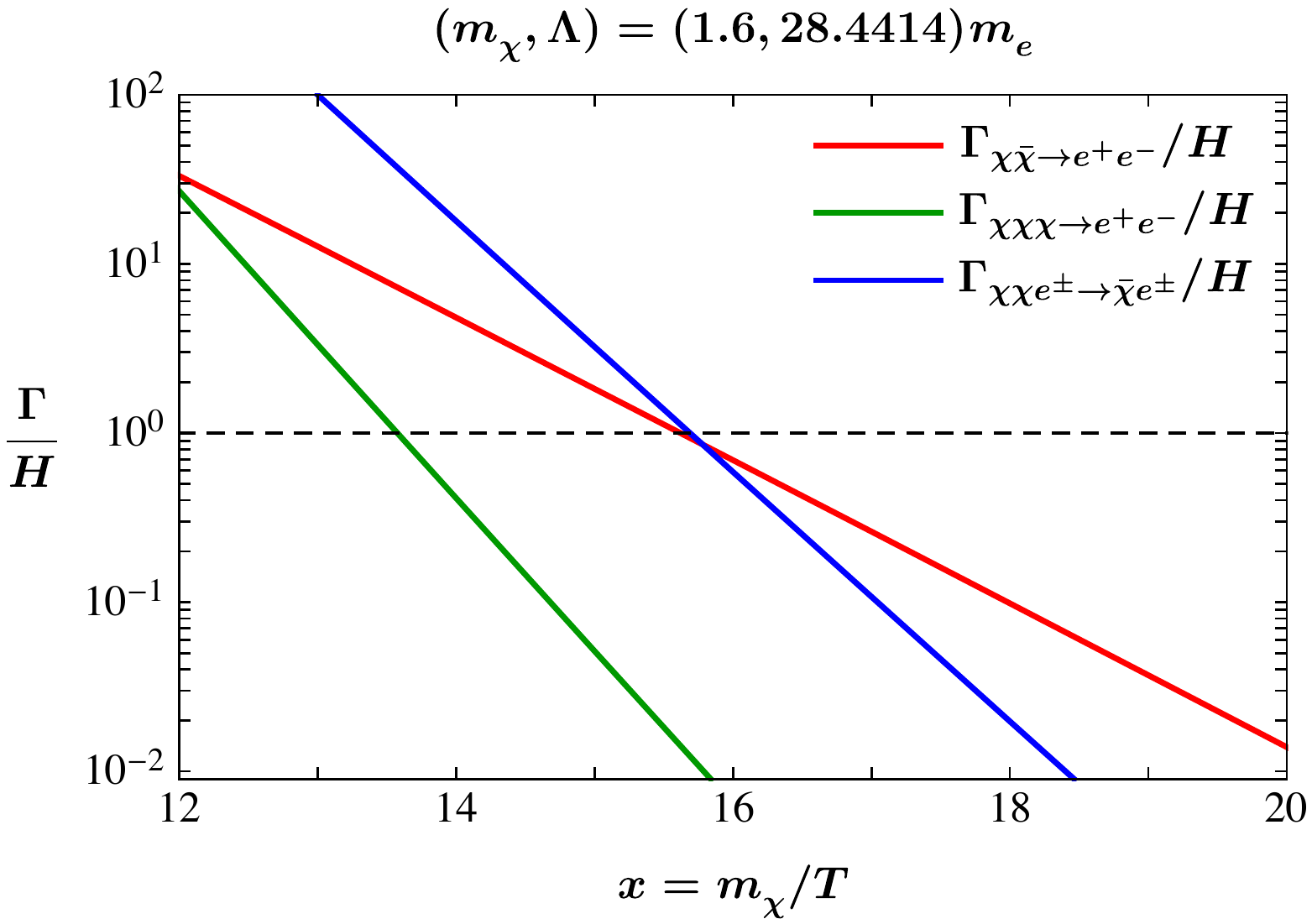}
\\[0.2cm]
\includegraphics[width=0.46\textwidth]{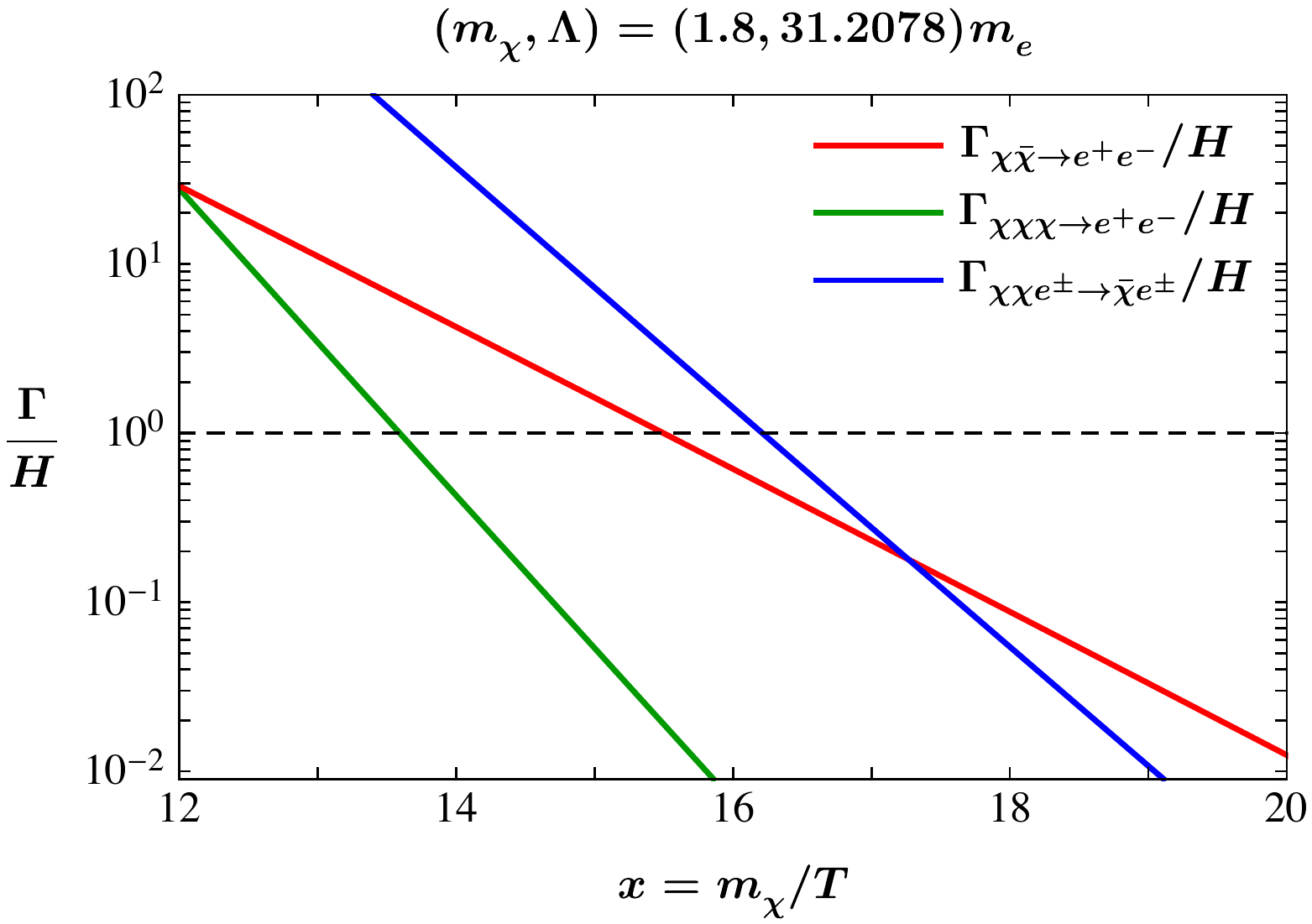}
\vs{-0.2cm}
\caption{$\Gamma/H$ around the $x^{}_\tf{f.o.}$ for $m^{}_e < m^{}_\chi < 2{}^{}m^{}_e$.
\label{fig:GammaH}}
\end{center}
\end{figure}

Lastly, let us compare the reaction rates of the $3 \to 2$ and $2 \to 2$ processes around the freezeout temperature, the definitions of them are given as follows\,\cite{Kolb:1990vq}
\begin{eqnarray}
\Gamma^{}_{\chi\bar{\chi} \to e^+e^-} 
&{}^{}{}^{}\equiv{}^{}{}^{}&
4{}^{}{}^{}\langle \sigma v \rangle^{}_{\hs{-0.03cm}\chi\bar{\chi} \to e^+ e^-} n^\tf{eq}_\chi
~,
\\
\Gamma^{}_{\chi\chi\chi \to e^+e^-} 
&{}^{}{}^{}\equiv{}^{}{}^{}&
12{}^{}{}^{}\langle \sigma v^2 \rangle^{}_{\hs{-0.03cm}\chi\chi\chi \to e^+ e^-} \big(n^\tf{eq}_\chi\big)\raisebox{0.1pt}{$^{\hs{-0.03cm}2}$}
~,
\\
\Gamma^{}_{\chi\chi e^\pm \to \bar{\chi} e^\pm} 
&{}^{}{}^{}\equiv{}^{}{}^{}&
2{}^{}{}^{}\langle \sigma v^2 \rangle^{}_{\hs{-0.03cm}\chi\chi e^\pm \to \bar{\chi} e^\pm} 
n^\tf{eq}_\chi {}^{} n^\tf{eq}_e ~,
\end{eqnarray}
\newline
where $n^\tf{eq}_i = s {}^{} Y^\tf{eq}_i$.\,\,We show in Fig.\,\ref{fig:GammaH} the ratios of the reaction rates to the Hubble expansion rate as functions of $x$ around the freezeout temperature with the parameter inputs given in Figs.\,\ref{fig:Ychi1} and \ref{fig:Ychi2}.\,\,Accordingly, the reaction rate of the $2 \to 2$ process is dominated within the mass range $1 \lesssim m^{}_\chi/m^{}_e \lesssim 1.6$, and is subdominated inside the mass range $1.6 \lesssim m^{}_\chi/m^{}_e \lesssim 2$. 
\vs{0.1cm}
 
\begin{thebibliography}{99}

%\cite{Smirnov:2020zwf}
\bibitem{Smirnov:2020zwf}
J.~Smirnov and J.~F.~Beacom,
%``New Freezeout Mechanism for Strongly Interacting Dark Matter,''
Phys. Rev. Lett. \textbf{125}, no.13, 131301 (2020).
%doi:10.1103/PhysRevLett.125.131301
%[arXiv:2002.04038 [hep-ph]].
%48 citations counted in INSPIRE as of 08 Aug 2021

%\cite{Ho:2021ojb}
\bibitem{Ho:2021ojb}
S.~Y.~Ho, P.~Ko and C.~T.~Lu,
%``Reshuffled SIMP Dark Matter,''
[arXiv:2107.04375 [hep-ph]].
%0 citations counted in INSPIRE as of 08 Aug 2021

%\cite{Gondolo:1990dk}
\bibitem{Gondolo:1990dk}
P.~Gondolo and G.~Gelmini,
%``Cosmic abundances of stable particles: Improved analysis,''
Nucl. Phys. B \textbf{360}, 145-179 (1991).
%doi:10.1016/0550-3213(91)90438-4
%1227 citations counted in INSPIRE as of 09 Aug 2021

%\cite{Saikawa:2018rcs}
\bibitem{Saikawa:2018rcs}
K.~Saikawa and S.~Shirai,
%``Primordial gravitational waves, precisely: The role of thermodynamics in the Standard Model,''
JCAP \textbf{05}, 035 (2018)
%%doi:10.1088/1475-7516/2018/05/035
%[arXiv:1803.01038 [hep-ph]].
%56 citations counted in INSPIRE as of 14 Jun 2021

%\cite{Kolb:1990vq}
\bibitem{Kolb:1990vq}
E.~W.~Kolb and M.~S.~Turner,
%``The Early Universe,''
Front. Phys. \textbf{69}, 1-547 (1990).
%2008 citations counted in INSPIRE as of 15 Jun 2021

%\cite{Bhattacharya:2019mmy}
\bibitem{Bhattacharya:2019mmy}
S.~Bhattacharya, P.~Ghosh and S.~Verma,
%``SIMPler realisation of Scalar Dark Matter,''
JCAP 01, 040 (2020).
%doi:10.1088/1475-7516/2020/01/040
%[arXiv:1904.07562 [hep-ph]].
%7 citations counted in INSPIRE as of 15 May 2021

%\cite{Aghanim:2018eyx}
\bibitem{Aghanim:2018eyx}
N.~Aghanim \textit{et al.} [Planck],
%``Planck 2018 results. VI. Cosmological parameters,''
Astron. Astrophys. 641, A6 (2020).
%doi:10.1051/0004-6361/201833910
%[arXiv:1807.06209 [astro-ph.CO]].
%5038 citations counted in INSPIRE as of 15 May 2021

\end{thebibliography}
\end{document}